\documentclass[natbib]{svjour3}                     
\smartqed  
\usepackage{graphicx}
\usepackage{bm,amssymb,amsmath,color,url,hyperref}
\newcommand{\mproton}{{\rm m_p}}    
\newcommand{\kB}{{\rm k_B}}         
\newcommand{\sigmaT}{\sigma_{\rm T}}         
\newcommand{\sigmaSB}{\sigma_{\rm SB}}         
\newcommand{\Msun}{{\rm M}_\odot}         
\newcommand{\Rsun}{{\rm R}_\odot}         
\newcommand{\D}{{\rm d}}         

\begin{document}

\title{Menus for Feeding Black Holes
}


\author{Bence Kocsis  \and Abraham Loeb
}


\institute{Bence Kocsis \at
               Institute for Advanced Study, Princeton, NJ, USA\\
              \email{bkocsis@ias.edu}           
           \and
           Abraham Loeb \at
              Harvard-Smithsonian Center for Astrophysics, Cambridge, MA, USA\\
              \email{aloeb@cfa.harvard.edu}           
}

\date{Received: 2 April 2013 / Accepted: 12 August 2013}

\maketitle

\begin{abstract}
Black holes are the ultimate prisons of the Universe, regions of
spacetime where the enormous gravity prohibits matter or even light to
escape to infinity.  Yet, matter falling toward the black holes may
shine spectacularly, generating the strongest source of radiation.
These sources provide us with astrophysical laboratories of extreme
physical conditions that cannot be realized on Earth.  This chapter
offers a review of the basic menus for feeding matter onto black holes and
discusses their observational implications.

\end{abstract}

\section{Introduction}
\label{s:intro}
There are several avenues for feeding matter onto black holes. Black
holes can accrete ambient gas from the interstellar medium or from an
attached gaseous disk, they may be fed by winds of a binary companion
star or directly accrete from a binary companion through Roche lobe
overflow, they can directly swallow or tidally disrupt stellar objects
that approach them, or they may grow by merging with other black
holes. Here we review these possibilities in detail and describe the
corresponding astrophysical models of accretion.

We start with an overview of the standard formation and evolutionary
scenarios of the three main classes of black holes: stellar mass,
intermediate mass, and supermassive black holes. In subsequent
sections we discuss the main classes of accretion including spherical
accretion, disk accretion, and accretion due to tidal disruptions of
stars.

\section{Black Hole Formation and Evolution}

\subsection{Collapse of Massive stars}

\begin{figure*}
\begin{center}
\includegraphics[angle=-90,scale=0.36]{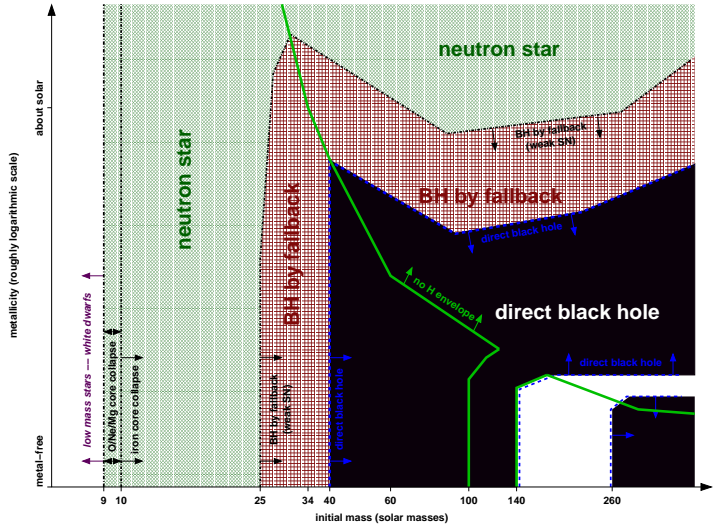}
\caption{
Types of stellar remnants of massive stars
for different initial stellar masses and metallicity.
 Figure credit: \citet{2003ApJ...591..288H}. }
\label{f:formation-collapse}
\end{center}
\end{figure*}

Compact objects may form when the core of a massive star consumes its
nuclear fuel and loses pressure support, resulting in a catastrophic
collapse. Whether the final state is a neutron star or a black hole depends 
on the mass of the star, its composition (i.e. metallicity) 
as depicted in Figure~\ref{f:formation-collapse}, and rotation speed. 
Stars with masses between $\sim 8$--$25$ often explode as a Type II 
supernova and leave a neutron star/pulsar remnant. For more massive stars, 
in the range $25$--$35\,\Msun$, the fall back of the carbon-oxygen core 
may be significant to crush the remnant into a black hole 
\citep{1999ApJ...522..413F}. For even more massive stars, the complete 
hydrogen envelope is lost to strong winds, leaving an isolated Wolf-Rayet star. 
This star then explodes as a Type-Ib/c supernova. Depending on the 
less-understood Wolf-Rayet mass loss rates, the final remnant is either 
a neutron star or a black hole. Recent work \citep[e.g.,][]{2012ApJ...757...69U} 
shows that black holes can  form across a somewhat wider range 
of initial masses ($M\gtrsim 15\,\Msun$), and neutron stars can form 
from much higher mass progenitors ($M\gg 25\,\Msun$). This is also 
hinted by the young progenitors inferred for observed magnetars 
($M > 40\,\Msun$; \citealt{2006ApJ...636L..41M}).

The mass of the remnant black hole depends on the amount of wind loss,
fall back, and metallicity.  It is typically estimated to be between
$5$--$15\,\Msun$ for solar metallicities and between $5$--$80\,\Msun$
in globular clusters or metal-poor galaxies with $0.01\,Z_{\odot}$
\citep{2010ApJ...714.1217B}. Higher mass black holes may form during
the collapse of the first stars, the so-called Population III stars,
discussed in \S~\ref{s:supermassive-stars}.

\subsection{Black Holes in Stellar Binaries}\label{s:binary}
\begin{figure*}
\begin{center}
\includegraphics[width=12cm]{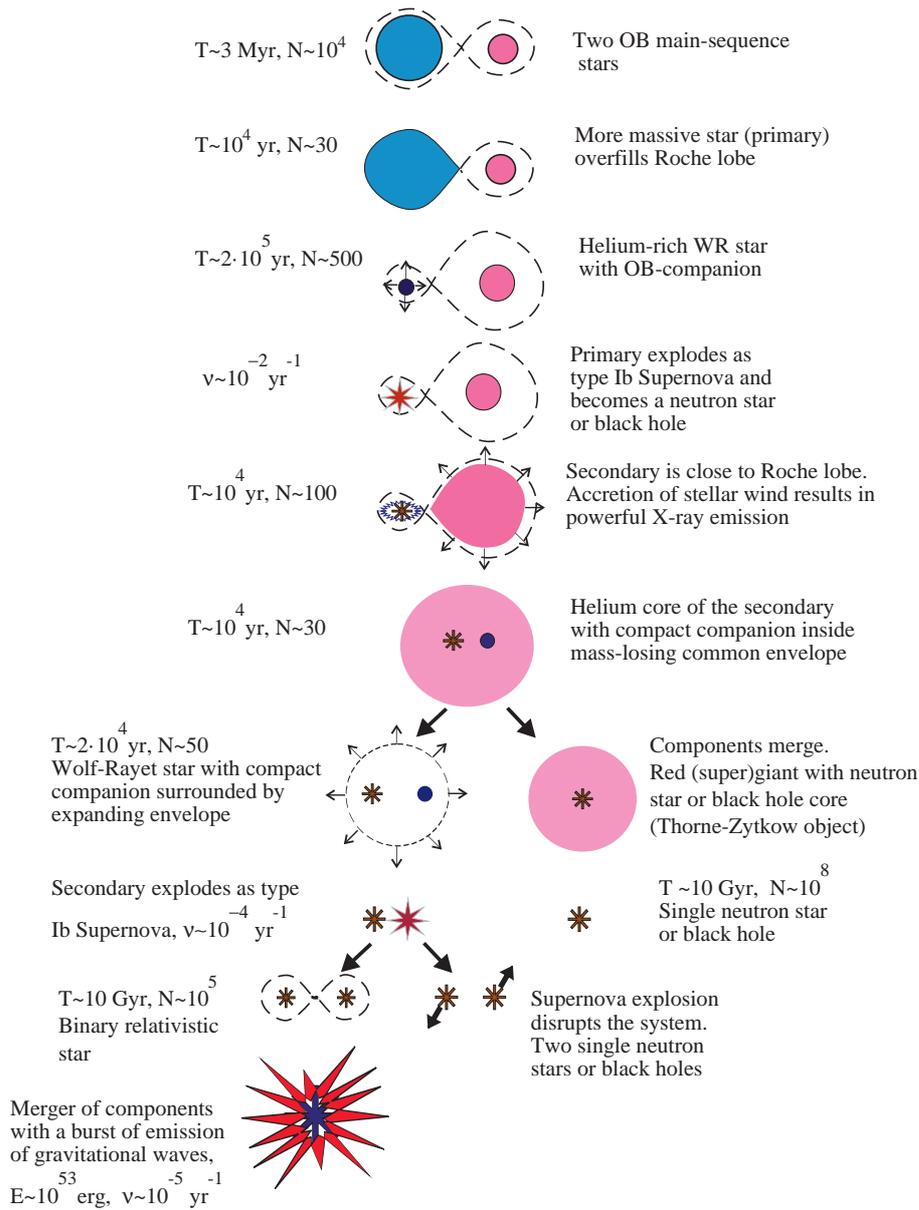}
\caption{\label{f:formation-hmxb} Evolutionary stages of high mass
binary stars from top to bottom.
The initially more massive star
(primary) is shown on the left.  A black hole or neutron star forms in
the $4^{\rm th}$ stage during a type Ib supernova possibly accompanied
by a long GRB. The $5^{\rm th}$ stage represents a typical HMXB.  Once
the secondary fills its Roche lobe, it may supply super-Eddington
accretion and lead to a super-soft X-ray source.  Later the binary
inspirals in a common envelope, ejecting the envelope.  This may lead
to a second supernova, leaving a relativistic double BH, NS, or BH/NS
binary. These objects merge after gravitational wave emission,
possibly generating a short GRB.
The approximate timescale and number of such binaries in the Galaxy
are labeled in each evolutionary stage.
Figure credit:
\citet{2006LRR.....9....6P}. }
\label{QLF}
\end{center}
\end{figure*}

Many stellar-mass black holes and neutron stars are found in close
binary systems.  In the local Universe, black-hole X-ray binaries come
in two flavors, depending on the mass of the companion star: {\it
low-mass X-ray binaries} where a low-mass companion transfers mass
owing to the tidal force exerted by the black hole, and {\it high-mass
X-ray binaries (BH-HMXB)} where the companion is a
massive star which could also transfer mass to the black hole through
a wind.  At redshifts $z\gtrsim 6$ when the age of the Universe was
short, BH-HMXB were probably most important since they are known to
produce their X-rays over a short lifetime ($\sim 10^8$ yr). The
cumulative X-ray emission from BH-HMXBs is expected to be proportional
to the star formation rate \citep{2003MNRAS.339..793G,2003A&A...399...39R,2012MNRAS.419.2095M}. 
If indeed the early population of stars
was tilted towards high masses and binaries were common, BH-HMXB may
have been more abundant per star formation rate in high redshift
galaxies.  The X-rays produced by BH-HMXBs may have had important
observable effects as they catalyzed H$_2$ formation, heated the
intergalactic medium, and modified the 21-cm signal from neutral
hydrogen. Their overall influence was, however, limited: hydrogen
could not have been reionized by X-ray sources based on current limits
on the unresolved component of the X-ray background.

The formation of BH binaries may be linked to the evolution of binary
stars, transferring mass during their lives.  X-ray binaries in
particular, are believed to be affected by mass transfer in the giant
phase.  The standard evolutionary scenario leading to a HMXB is
depicted in Fig.~\ref{f:formation-hmxb}. Here at least one of the
stars in the binary must be a massive star.  After a few million
years, it exhausts its hydrogen, expands into a red giant, and
transfers its entire envelope to the secondary. At this stage, the
primary and secondary become a Wolf-Rayet star and a massive O/B
star. The primary then explodes as a type Ib supernova, and becomes a
neutron star or a black hole depending on its initial mass (see
Fig.~\ref{f:formation-collapse}).  The powerful winds of the massive
secondary star sources accretion for the companion compact
object. Later, as the secondary runs out of hydrogen and expands to
within the  {\it Roche lobe} of the compact object 
where the gravity of the compact object dominates, 
\begin{equation}\label{e:Roche}
r_R = a(1-e) \left( \frac{m}{2 M} \right)^{1/3}
\end{equation}
the compact object will accrete
continuously from the secondary. Here $M$, $m$, $a$, $e$ are the
primary and secondary masses, semimajor axis, and eccentricity.  As
the secondary becomes a red giant, it may completely engulf the
orbit. In this common envelope stage, the angular momentum of the
binary can quickly decrease as the envelope is heated and ejected.  If
the core of the giant merges with the neutron star, it may form a
Thorne-Zytkow object \citep{1977ApJ...212..832T},
or if it is ejected completely, it leaves behind a black hole or
neutron star primary orbiting closely around a Wolf-Rayet star
secondary. Next, the secondary explodes as a Type-Ib supernova.  If
the kick from the supernova explosion is larger than the gravitational
binding energy, the two compact objects may fly apart, as single black
holes or neutron stars.  Alternatively, if the kick is weaker, it
leaves behind a relativistic eccentric double black hole, double
neutron star, or black hole-neutron star binary.

These double compact object binaries then gradually loose energy and angular momentum
due to gravitational wave emission, which leads to the shrinking of the orbital
separation and eccentricity.
The existence of gravitational waves is a generic prediction of Einstein's theory of
gravity. They represent ripples in space-time generated by the motion
of the two black holes as they move around their common center of mass
in a tight binary.  The energy carried by the waves is taken away from
the kinetic energy of the binary, which therefore tightens with time.
Ultimately, the gravitational waves emitted close to merger
will be detectable by Advanced LIGO\footnote{\url{http://www.ligo.caltech.edu/}}
and VIRGO\footnote{\url{http://www.ego-gw.it/}}.
Additionally, if at least one of the components is a neutron star, the
merger may produce a bright gamma ray burst (GRB).

The above scenario involves massive stars. The analogous evolutionary
scenario for less massive binary stars leads to a white dwarf, neutron
star, or black hole with a moderate to low-mass main sequence star,
and ultimately a double compact object binary (e.g. double degenerate
white dwarf binary).  In this case, the separation is much larger
larger than commonly observed in low mass X-ray binaries (LMXBs). The
formation of LMXBs is less understood.  Many-body interactions,
especially in globular clusters may lead to LMXBs. Alternatively, a
massive binary system may form LMXBs if there is a rapid
super-Eddington wind phase and angular momentum evolution in between
the $5^{\rm th}$ and $6^{\rm th}$ stage in
Figure~\ref{f:formation-hmxb}. 
Black holes may form in low mass binaries during an accretion-induced collapse
if a neutron star is feeding through Roche lobe overflow from a companion  
\citep{2006ApJ...643L..13D}. Another possibility is the merger of a WD
with a NS \citep{2009PhRvD..80b4006P}, 
although in this case it is unclear whether sufficient mass will accrete onto the NS 
if the WD material undergoes explosive nuclear burning \citep{2013ApJ...763..108F}.

Neutron star or black hole binaries that accrete from a strong wind of
the companion star, may be modeled using Bondi-Hoyle-Lyttleton
accretion.  Compact objects that accrete from the companion star
through a Roche-lobe overflow, provide examples of disk accretion. We
discuss Bondi and disk accretion in \S~\ref{s:bondi} and
\S~\ref{s:disk}, respectively.

A neutron star in a LMXB that undergoes mass transfer from a secondary
star is expected to be spun up during the accretion to millisecond
periods, leaving a millisecond pulsar. Since black holes are more
massive with a larger angular momentum, they cannot be spun up
significantly by a lower mass companion.

\subsection{Supermassive black holes}
\label{s:smbh}

Supermassive black holes of masses between $10^{6}$--$10^{10}\,\Msun$ are
observed in the centers of most if not all nearby galaxies.
As they accrete gas, the gas shines as a bright point-like quasi-stellar object,
called a quasar. These objects are observed from cosmological distances
up to high redshifts, where the Universe was less than a billion years old.
How could such massive black holes form in such a short amount of time?

Growing a supermassive black hole very quickly is difficult.
Accretion of collisionless dark matter particles is negligible and can
be ignored \citep{Collisionless}.
For gas accretion -- there is a maximum luminosity at which the
environment of a black hole of mass $M_{\rm BH}$ may shine and still
accrete gas, called the Eddington limit, $L_{\rm Edd}$.  This limit is
obtained by setting the outward continuum radiation pressure equal to
the inward gravitational force. Denoting the gravitational potential
with $\Phi$, pressure with $p$, density with $\rho$,
\begin{equation}\label{e:radiationeq}
 \nabla \Phi = -\frac{\nabla p}{\rho} = \frac{\kappa}{c} \bm{F}_{\rm rad}\,,
\end{equation}
where in the last equality we assumed that the pressure is dominated by
radiation pressure which is associated to a radiation flux $\bm{F}_{\rm rad}$.
Here $\kappa$, is the opacity.
There are two primary sources of opacity for the typical densities and temperatures
here: Thomson electron scattering and bremsstrahlung
(i.e labeled free-free absorption) with
\begin{equation}\label{e:kappa}
\kappa_{\rm es}=\frac{\sigmaT}{\mproton}=0.4~{\rm cm}^2\,{\rm g}^{-1}\quad{\rm and}~~
\kappa_{\rm ff}
\approx 8\times 10^{22} {\rm cm}^2\,{\rm g}^{-1}
\left(\frac{\rho}{{\rm g}\,{\rm cm}^{-3}}\right) \left(\frac{T}{\rm K}\right)^{-7/2}\,,
\end{equation}
where we assumed a pure hydrogen plasma for simplicity, where
$\mproton$ denotes proton mass and $\sigmaT$ denotes the Thomson cross
section.  Substituting equation~(\ref{e:radiationeq}) into the
definition of the luminosity of the source bounded by a surface $S$,
we get
\begin{equation}
 L = \int_S \bm{F}_{\rm rad}\cdot \bm{\D S} = \frac{c}{\kappa} \int_S \nabla \Phi \cdot \bm{\D S}\,.
\end{equation}
Using Gauss's theorem, Poisson's equation $\nabla^2\Phi = 4\pi G \rho$, and
the definition of the mass,
\begin{equation}\label{e:LE}
 L_{\rm E} =  \frac{c}{\kappa} \int_V \nabla^2 \Phi \D V = \frac{4\pi G c}{\kappa} \int_V \rho \D V = \frac{4\pi G M c}{\kappa}
= 1.3\times 10^{44} \left({M_{\rm BH} \over
10^6 \ M_\odot}\right)\,{\rm erg\,s^{-1}}\,.
\end{equation}
This sets the maximum luminosity of a source in hydrostatic equilibrium,
$L_{\rm Edd}$ denotes Eddington luminosity.
Note that we did not assume spherical geometry. While the effects of rotation
can somewhat change this theoretical limit in an accretion disk,
observed quasars for which black hole masses
can be measured by independent methods appear to respect this limit.

The total  luminosity from gas
accreting onto a black hole, $L$, can be written as
some radiative efficiency $\epsilon$ times the mass  accretion rate $\dot{M}$,
\begin{equation}\label{e:efficiency}
L=\epsilon \dot{M} c^2,
\end{equation}
with the  black hole accreting the non-radiated
component, $\dot{M}_{\rm BH}= (1-\epsilon)\dot{M}$. The equation that
governs the growth of the  black hole mass is then
\begin{equation}
\dot{M}_{\rm BH}={M_{\rm BH}\over t_{\rm Edd}},
\end{equation}
where (after substituting all fundamental constants),
\begin{equation}
t_{\rm Edd}= 4\times 10^7 \left({\epsilon/(1-\epsilon)\over
0.1}\right)\left({L\over L_{\rm Edd}}\right)^{-1} \ {\rm yr}.
\label{tE}
\end{equation}
We therefore find that as long as fuel is amply supplied, the black
hole mass grows exponentially in time, $M_{\rm BH}\propto \exp
(t/t_{\rm Edd} )$, with an $e$-folding time $t_{\rm Edd}$. Since the
growth time in equation (\ref{tE}) is significantly shorter than the
$\sim 10^9$ years corresponding to the age of the Universe at a
redshift $z\sim 6$ -- where black holes with a mass $\sim 10^9 \
M_\odot$ are found -- one might naively conclude that there is plenty
of time to grow the observed black hole masses from small seeds.  For
example, a seed black hole from a Population III star of $100 \
M_\odot$ can grow in less than a billion years up to $\sim 10^9\
M_\odot$ for $\epsilon \sim 0.1$ and $L\sim L_{\rm Edd}$.  However,
the intervention of various processes makes it unlikely that a stellar
mass seed will be able to accrete continuously at its Eddington limit
without interruption.

Mergers are very common in the early Universe.  Every
time two gas-rich galaxies come together, their black holes are likely
to coalesce. The coalescence is initially triggered by ``dynamical
friction'' from the surrounding gas and stars (see Fig.~\ref{s:friction}, and is completed --
when the binary gets tight -- as a result of the emission of
gravitational radiation \citep{1980Natur.287..307B}.
Computer simulations reveal that when two black holes with unequal
masses merge to make a single black hole, the remnant gets a kick due
to the non-isotropic emission of gravitational radiation at the final
plunge.  This kick was calculated recently
using advanced computer codes that solve Einstein's equations, a task
that was plagued for decades with numerical
instabilities \citep{2005PhRvL..95l1101P,2006PhRvL..96k1101C,2006PhRvL..96k1102B}.
The typical kick velocity is hundreds of
kilometer per second (and up to ten times more for special spin
orientations), much larger than the escape speed from the first dwarf
galaxies. This implies that continuous accretion was likely punctuated
by black hole ejection events \citep{2008MNRAS.390.1311B,2009ApJ...696.1798T},
forcing the merged dwarf galaxy to grow a new black hole seed from
scratch.
These black hole recoils might have left observable
signatures in the local Universe.  For example, the halo of the Milky
Way galaxy may include hundreds of freely-floating ejected black holes
with compact star clusters around them, representing relics of the
early mergers that assembled the Milky Way out of its original
building blocks of dwarf galaxies \citep{2009MNRAS.395..781O}.

\subsection{Seeds of Supermassive Black Holes} \label{s:supermassive-stars}
Supermassive black holes with masses exceeding $10^9\,\Msun$ are
observed at cosmological distances where the Universe was less than a
billion years old.  Assuming that their luminosity does not exceed the
Eddington limit and the radiative efficiency is $\sim 10\%$, one is
driven to the conclusion that the black hole seeds must have started
more massive than the stellar regime of $\lesssim 100 \ M_\odot$.  The
needed seeds may originate from \emph{supermassive stars}, defined as
hydrostatic configurations with masses $10^3$--$10^8 \ M_\odot$
\citep{1994ApJ...432...52L,2004ARA&A..42...79B}.
Lacking carbon, nitrogen, oxygen, and iron, these first stars do not
drive powerful stellar winds as present-day massive stars.  Such
systems have not been observed as of yet. Theoretically, they are
expected to be supported almost entirely by radiation pressure and
hence their luminosity equals the Eddington limit
(equation~\ref{e:LE}).  Supermassive stars steadily contract and
convert their gravitational binding energy to radiation with a total
lifetime $\lesssim 10^6$~yr before they collapse to a black hole.

{\it Is it possible to make such stars in early galaxies?}
Numerical simulations indicate that stars of mass $\sim 10^6 \
M_\odot$ could have formed at the centers of early dwarf galaxies that
were barely able to cool their gas through transitions of atomic
hydrogen, having $T_{\rm vir}\sim 10^4$~K and no H$_2$ molecules
(which were suppressed by a Lyman-Werner background,
\citealt{2003ApJ...596...34B,2008MNRAS.391.1961D,2009MNRAS.396..343R}).
Such systems have a total mass that is several orders of magnitude
higher than the earliest Jeans-mass condensations.  In both cases, the
gas lacks the ability to cool well below $T_{\rm vir}$, and so it
fragments into one or two major clumps.  The simulation results in
clumps of several million solar masses, which inevitably form massive
black holes.  The existence of such massive seeds would have given a
jump start to the black hole growth process.

The nearly uniform entropy established by convection makes the
structure of supermassive stars simple (equivalent to a so-called
polytrope with an index $n=3$) with a unique relation between their
central temperature $T_c$ and central density $\rho_c$,
\begin{equation}
T_c=2\times 10^6~{\rm K}\left({\rho_c\over
1~{\rm g~cm}^{-3}}\right)^{1/3}\left({M_\star\over 10^6 \ M_\odot}\right)^{1/6}.
\end{equation}
Because of this modest temperature, nuclear reactions are insignificant in metal-poor stars with
masses $M_\star>10^5 \ M_\odot$.  General relativistic corrections make
the star unstable to direct collapse to a black hole as soon as its
radius contracts to a value
\begin{equation}
R_\star<R_{\rm crit}=1.59\times 10^3 \left({M_\star\over
10^6 \ M_\odot}\right)^{1/2} \left({GM_\star\over c^2}\right) .
\end{equation}
Rotation can stabilize supermassive stars to smaller radii, but even
rotating stars are expected to eventually collapse to a black hole
after shedding their angular momentum through a wind \citep{2002ApJ...569..349S}.
If the supermassive star is made of pre-enriched gas, then powerful
winds will inevitably be driven at its surface where the opacity due
to lines from heavy elements far exceeds the Thomson value, making the
outward radiation force stronger than gravity.

We note that the infall of a sufficiently dense, optically-thick
spherical envelope of gas cannot be prevented by radiation pressure
even if the radiation production rate exceeds the Eddington limit near
the center. So long as the mass infall rate is sufficiently high, the
Eddington limit will not apply because the photons will be trapped in
the flow.  Super-Eddington accretion can grow a seed black hole
rapidly (as in the case of stellar collapse), as long as the blanket
of infalling gas advects the radiation inwards as it accretes onto the
black hole.  This ``obscured'' mode of black hole accretion (which is
hidden from view) could be particularly important at high redshifts
when the gas density and infall rate onto galaxies obtain their
highest values \citep{2012MNRAS.425.2892W}.
We discuss the physics of this possibility in \S~\ref{s:bondi} below.

\subsection{Intermediate mass black holes} \label{s:imbh}
Up to this point we have discussed the two well known classes of black
holes: the stellar-mass and supermassive black holes. An intriguing
question is whether there is an intermediate population of black holes
in the mass range $M_{\rm imbh}\sim100$--$10^5\Msun$
\citep{2004IJMPD..13....1M}. These objects, referred to as
intermediate mass black holes (IMBHs), are interesting for several
reasons: they may represent seeds for forming SMBHs through accretion,
they could stabilize globular clusters against core collapse
\citep{2004ApJ...613.1143B}, they can lead to dark matter
overdensities which will cause excessive dark matter annihilation
signals \citep{2005PhRvD..72j3517B,2009NJPh...11j5016B}, they may have
participated in cosmic reionization \citep{2004ApJ...604..484M}, and
may provide sources of gravitational waves for direct detection
\citep{2003ApJ...590..691W,2004ApJ...612..597W,2008ApJ...681.1431M,2009MNRAS.395.2127O,2012ApJ...752...67K}.

Several theoretical arguments have been put forth for forming IMBHs.
The collapse of the first supermassive Pop-III stars may form IMBHs. This process can add
$\sim 50(M_{\rm imbh}/150\,\Msun)$ IMBHs to galactic centers \citep{2001ApJ...551L..27M}.
Secondly, runaway collisions of stars \citep{2002ApJ...576..899P,2006MNRAS.368..141F}
or collisions with stellar black holes \citep{2006ApJ...637..937O}
in dense star clusters can also produce IMBHs. The clusters
sink to the galactic nucleus due to dynamical friction (\S~\ref{s:friction}) and deposit
their IMBHs. This channel may supply an additional 50 IMBHs of mass $10^3\,\Msun$
within the inner 10\,pc \citep{2006ApJ...641..319P}.
IMBHs can also form and migrate inwards in the
accretion disks of SMBHs \citep{2004ApJ...608..108G,2007MNRAS.374..515L,2011PhRvD..84b4032K,2012MNRAS.425..460M}.

Ultraluminous X-ray sources (ULXs) provide the best observational
candidates for IMBHs (see \S~\ref{s:disk} below).  In particular, the
estimated IMBH mass in ESO 243-49 HLX-1 based on its X-ray thermal
accretion disk spectrum and radio signal is in the range $9\times
10^3$--$9\times 10^4\,\Msun$
\citep{2011ApJ...734..111D,2012Sci...337..554W}.  IMBHs may also
represent low luminosity X-ray sources in the Galactic bulge accreting
gas from the interstellar medium or infalling gas clouds
\citep{2013arXiv1302.3220B}.

\section{Bondi-Hoyle-Lyttleton Accretion}\label{s:bondi}
Bondi-Hoyle-Lyttleton accretion denotes a general class of black hole
accretion, where the black hole is completely embedded in a gaseous
medium and the inflow is spherical without a significant
rotation. This occurs if black holes accrete from the ambient
interstellar medium, as they pass through a dense gas cloud, become
embedded in the envelope of a giant star, or if they accrete from a
powerful wind of a binary companion star. Stellar mass black holes
embedded in the accretion disk of a supermassive black hole may also
accrete from the gaseous medium through this mode of accretion. Here
we review the physical principles of Bondi-Hoyle-Lyttleton accretion
and its observational implications.

\subsection{Simple model for spherical accretion}
Consider a black hole of mass $M$ is moving with relative velocity $V$ in an ambient
medium of density $\rho_0$ and temperature $T_0$.
The root-mean-square velocity of thermal protons in the gas relative to the black hole is roughly
$v_{\rm rms}=\sqrt{c_s^2 + V^2}$, where $c_s\approx {\sqrt{\kB T / \mproton}}$
is the sound speed .
The gas particles interior to the Bondi radius,
\begin{equation}\label{e:rBondi}
r_{\rm B} =  \frac{2 G M}{ v_{\rm rms}^2 }
\end{equation}
are gravitationally bound to the black hole and are accreted.  The
steady mass flux of particles entering this radius is $\rho_0 v_{\rm
rms}$.  Multiplying this flux by the surface area associated with the
Bondi radius, $\pi r_{\rm B}^2$, gives the supply rate of gas,
\begin{equation}
\dot{M}_{\rm B}=
4\pi\rho_0 \frac{G^2 M^2}{v_{\rm rms}^3} =78 \left({M\over 10^8 \ M_\odot}\right)^2 \left({n_H\over 1~{\rm
cm^{-3}}}\right) \left({T_0\over 10^4~{\rm
K}}\right)^{-3/2}~{M_\odot~{\rm yr}^{-1}} .
\label{e:MBondi}
\end{equation}
where the second equality assumes a static medium ($v_{\rm rms}=c_s$)
and $n_{H} = \rho_0/\mproton$. In a steady state this supply rate equals the mass
accretion rate into the black hole.

This simple estimate for quasi-spherical accretion onto black holes is
consistent with a range of simplified models.
\citet{1939PCPS...35..405H} derived the accretion onto a point mass in
the limit that the gas pressure is negligible ($c_s=0$) and particles
move on ballistic orbits. \citet{1944MNRAS.104..273B} extended the
analysis to include accretion from an axisymmetric wake behind the
black hole, which led to a similar result, modified by a factor $\sim
1/2$.  \citet{1952MNRAS.112..195B} solved the Euler and continuity
equations for a spherically symmetric steady-state adiabatic flow of
gas assuming that the accreting object does not move relative to the
medium ($V=0$).  Well inside the sonic radius (i.e., the point at
which the infall and sound speeds are equal), the velocity is close to
free-fall $v\sim(2GM/r)^{1/2}$ and the gas density is $\rho(r)\sim
\rho_0 (r/r_{\rm B})^{-3/2}$. If the black hole moves with a
supersonic velocity $V>c_s$, then a shock wave forms behind the hole,
and the accretion occurs primarily from the shocked gas in a column
trailing the black hole.  The Bondi solution is then appropriate
interior to the shock front and the Bondi radius $r_B$.  The
correction factor of this detailed calculation for the accretion rate
relative to equation~(\ref{e:MBondi}) is of order unity (see
\citealt{1994ApJ...427..342R} for an updated formula).

\subsection{Luminosity}

The luminosity of an accretion flow is generally expressed in terms of
the radiative efficiency $\epsilon$ for converting rest mass to
radiation as, $ L = \epsilon \dot{M} c^2$.  The accretion rate for
Bondi-Hoyle-Lyttleton accretion, $\dot{M}_B$, is given by
equation~(\ref{e:MBondi}).  The radiative efficiency, $\epsilon$ is
very small during spherical accretion since most of the energy, in the
form of kinetic energy, heat, and radiation, is advected into the
black hole rather than escape to infinity.  This is due to the fact
that the cooling time is longer than its accretion (free-fall) time or
the diffusion time of the radiation outwards is much longer than the
free-fall time and the radiation is trapped in the flow
\citep{1978PhyS...17..193R,1979MNRAS.187..237B,1986ApJ...308..755B}.
The infall time of the fluid element from the Bondi radius to to the
BH is approximately, $t_{\rm ff}(r) = r_{\rm B}/v_{\rm rms}$ and the
photon diffusion time in the optically thick limit (i.e. where the
optical depth satisfies $\tau = \kappa \rho r_B \gg 1$) is
\begin{equation}\label{e:tdiff}
t_{\rm diff}(r) = \frac{1}{2c} \int_{0}^{r_{\rm B}} \kappa\rho(\xi)\xi \D \xi
= \frac{\kappa}{\pi c} \frac{\dot{M}_{\rm B}}{v_{\rm rms}}
\end{equation}
where we have used equations~(\ref{e:rBondi})
and (\ref{e:MBondi}), $\rho(r)\propto r^{-3/2}$ for Bondi accretion,
and $\kappa$ denotes the opacity.
Thus, the diffusion time is larger than the infall time if
\begin{align}\label{e:MBondicrit}
\dot{M}_{\rm B} \gtrsim
\frac{\pi r_{\rm B} c}{\kappa}
\simeq \frac{2\pi \,  G M c}{\kappa v_{\rm rms}^2} = \frac{\dot{M}_{\rm Edd}}{2}  \left(\frac{v_{\rm rms}}{c}\right)^{-2}\,.
\end{align}
where in the last equality we used the definition of the Eddington
luminosity, $L_{\rm Edd}$ and the Eddington accretion rate
$\dot{M}_{\rm Edd}=L_{\rm Edd}/c^2$.  The radiation is trapped in the
flow for hyper-Eddington accretion rates that satisfy
equation~(\ref{e:MBondicrit}).  For smaller $\dot{M}_{\rm B}$, the
inflowing gas can cool radiatively.  Radiation-hydrodynamical
simulations show that the flow remains radiatively inefficient, and
the luminosity significantly sub-Eddington even if the accretion is
super-Eddington.  For $300 \geq \dot{m}_{\rm B}\equiv \dot{M}_B/{\dot
M}_{\rm Edd}\gg1$, the radiative efficiency is estimated to be in the
range between $10^{-6}\lesssim \epsilon \lesssim 10^{-3}$
\citep{2012ApJS..201....9F} and $\epsilon\sim10^{-2}$
\citep{2012MNRAS.426.1613R}.  If the accretion rate is very
sub-Eddington, $\dot{m}_{\rm B} \ll 0.1$,
the radiative efficiency is expected to follow $\epsilon =10
\dot{m}_{\rm B} $ (see \S~\ref{s:adaf} below).

\subsection{Dynamical Friction}\label{s:friction}
In addition to mass accretion, a gravitating object experiences a drag when moving
through a medium. The drag arises due to the gravitational focusing of material
forming a wake behind the accretion. This process in collisionless
astronomical systems is called dynamical friction \citep{1943ApJ....97..255C,2008gady.book.....B}.
It is responsible for delivering massive black holes to the centers of galaxies after galaxy mergers
\citep{1980Natur.287..307B}, sinking satellites in dark matter galaxy halos and
thereby transporting stars and possibly intermediate mass black holes to the galactic nucleus
\citep{2006ApJ...641..319P}. A similar effect arises in a gaseous medium
\citep{1980ApJ...240...20R,1999ApJ...513..252O}.
The steady-state drag force is
\begin{equation}
 F_{\rm DF} = -4 \pi \rho_0 \frac{G M}{V^2} I
\end{equation}
where
\begin{align}
 I_{\rm supersonic} &= \ln\left( \frac{r_{\max}}{r_{\min}} \right) + \frac{1}{2} \ln\left( 1 - \frac{1}{{\cal M}^2} \right) ~~{\rm if}~~{\cal M}>1\,,\\
 I_{\rm subsonic} &= \frac{1}{2}\ln \left(\frac{1+ {\cal M} }{ 1- {\cal M} } \right) - {\cal M} ~~{\rm if}~~{\cal M}<1\,,
\end{align}
and ${\cal M} = V/c_s$ is the Mach number, $r_{\max}$ denotes the maximum size of the medium
and $r_{\min}$ denotes the size of the perturber. For black holes $r_{\min}$ is of order the gravitational radius, $G M/ c^2$.
In the limit ${\cal M}\gg 1$, $F_{\rm DF}\approx -\dot{M}_B V \ln(r_{\max}/r_{\min}) $,
and for ${\cal M}\ll 1$, $F_{\rm DF} \approx -\frac{1}{3}\dot{M}_B V $.

\subsection{Vorticity, turbulence, and radiation effects}
The simple model of Bondi-Hoyle-Lyttleton accretion neglects rotation,
gas cooling, radiative effects, turbulence, and relativistic effects,
which may modify the results significantly. These effects may be
studied using state-of-the-art numerical simulations, which is
becoming possible only recently.

\citet{2005ApJ...618..757K} examined Bondi accretion if the gas has a
non-zero angular momentum beyond the Bondi radius. Neglecting
radiation and thermodynamic effects, their simulation results are
consistent with the fitting formula
\begin{equation}\label{e:Krumholz}
 \dot{M}_{\rm vort} = \dot{M}_B \times
\left\{
\begin{array}{ll}
0.4 &  {\rm if~~}\omega_* \ll 1 \\
0.08\, \omega_*^{-1}\ln (16 \omega_*) &  {\rm if~~}\omega_* \gg 1\\
\end{array}
\right.
\end{equation}
where $\omega_* = \omega r_{\rm B}/(2c_s)$ is the vorticity parameter, where
$\omega$ is the angular velocity of the inflowing gas at the outer boundary with respect to the black hole.
This shows that the accretion rate is significantly suppressed by vorticity.

Further, \citet{2006ApJ...638..369K} have shown that the accretion
rate is significantly modified in a turbulent medium over the estimate
one obtains by using the turbulent velocity dispersion in
equation~(\ref{e:MBondi}).  In this case the accretion follows a
lognormal distribution with a median that is roughly given by
\begin{equation}
 \dot{M}_{\rm turb} = \left[ \dot{M}_{\rm B}^{-2} +  \dot{M}_{\rm vort}^{-2}  \right]^{-1/2}\,.
\end{equation}
These estimates assumed that the velocity of the object is much smaller than the
turbulent velocity dispersion.

Incorporating radiation and realistic heating and cooling processes
between multiple components (electrons, ions, and photons) poses a
challenge to realistic numerical simulations of Bondi-Hoyle-Lyttleton
accretion.  Recently, \citet{2012ApJS..201....9F} have examined
optically thick Bondi accretion using a general relativistic radiation
magneto-hydrodynamic (GR-R-MHD) simulation with $10 \leq
\dot{M}/\dot{M}_{\rm Edd}\leq 300$.  They assumed two components:
radiation and gas, with two gas-cooling mechanisms, Thomson scattering
and thermal bremsstrahlung. The ambient gas temperature was between
$10^5$ and $10^7\,{\rm K}$.  They found that the radiation pressure
remained sub-dominant throughout the flow, and the luminosity remained
very sub-Eddington despite the super-Eddington accretion rates such
that the radiative efficiency was $\epsilon=L/{\dot M c^2}\lesssim
10^{-4}$.  These simulations suggest that radiation does not have a
major effect on the results.  The opposite conclusion was reached
based an independent code by \citet{2011MNRAS.417.2899Z} and
\citet{2012MNRAS.426.1613R}.
These simulations also assumed two components (gas+radiation) with Thomson scattering and
bremsstrahlung. In these simulations, radiation pressure dominated over gas pressure, and
suppressed accretion by two orders of magnitude. The accretion rate was still super-Eddington in
the simulation, while the luminosity was around the Eddington limit with $\epsilon\sim 10^{-2}$.
These simulations showed that radiation pressure may serve to further stabilize the flow against
a flip-flop instability (see \S~\ref{s:bondi:instability}). While this progress is exciting,
further improvements are necessary for a more secure estimate of the radiative efficiency
of Bondi-Hoyle-Lyttleton accretion: primarily by incorporating the cooling effects of electrons
(synchrotron emission and Comptonization) and extending calculations to smaller temperatures
and higher resolution.

\subsection{Instabilities}\label{s:bondi:instability}
The Bondi-Hoyle-Lyttleton flow in not stable.  Two dimensional
numerical simulations have shown that the wake oscillates back and
forth, exhibiting a ``flip-flop instability''
\citep{1987A&A...176..235A,1988ApJ...335..862F,2009ApJ...700...95B,2013MNRAS.429.3144L},
however, the structure of the wake is more regular in 3D simulations
\citep{1999A&A...346..861R}.  At very high Mach numbers, the accretion
line behind the black hole is subject to a longitudinal
\citep{1977MNRAS.180..491C} and transverse instabilities
\citep{1990ApJ...358..545S,1991ApJ...376..750S}.  Furthermore,
\citet{1994MNRAS.270..871N} has shown that an an axisymmetric shock is
susceptible to a radial instability if the flow is accelerated after
the shock.  \citet{1999A&A...346..861R} investigated the
Kelvin-Helmholtz and Rayleigh-Taylor instabilities, and concluded that
these are not sufficient to explain the simulations without a feedback
mechanism.

Bondi-Hoyle-Lyttleton accretion in a supersonic medium
is subject to the standing accretion shock instability (SASI).
The physical reason for this instability is the advective-acoustic cycle
\citep{2002A&A...392..353F,2005A&A...435..397F,2012MNRAS.421..546G}:
entropy/vorticity perturbations are generated at the shock and
advected to the sonic point, where an acoustic wave is excited and
propagated back to the shock, leading to the growth of the entropy/vorticity
perturbation. SASI helps drive supernova explosions \citep{2006ApJ...640..878B}
and causes the emission of gravitational waves \citep{2007ApJ...655..406K}.

Recent three dimensional radiation-hydrodynamical simulations revealed
strong transient oscillations as the radiation pressure increases and
becomes comparable to the gas pressure, eventually reversing the shock
from the downstream to the upstream domain
\citep{2012MNRAS.426.1613R}.

\section{Disk Accretion} \label{s:disk}

\subsection{Thin Disk Accretion}
If the inflow is endowed with rotation with respect to the black hole,
it reaches a centrifugal barrier from where it cannot accrete farther
inwards unless its angular momentum is transported away. Near the
centrifugal barrier, where the gas is held against gravity by
rotation, an accretion disk forms around the black hole, centered
on the plane perpendicular to the rotation axis.  The accretion time
is dictated by the rate at which angular momentum is transported
through viscous stress which could be significantly longer than the
free-fall time for a non-rotating flow (as in the Bondi solution
described in \S~\ref{s:bondi}).  In the absence of radiative
processes, the dissipation of the gas's kinetic energy into heat would
make the disk thick and hot, with a kinetic energy of the gas close to
half the gravitational potential energy (virial equilibrium).  For
protons at distance $r$, this corresponds to a temperature $\sim
10^{13}~{\rm K}(r/r_{\rm g})^{-1}$ where $r_{\rm g}=GM/c^2$ is the
gravitational radius.  However, if the cooling time of the gas is
shorter than the viscous time, then a thin disk would form. This
latter regime is realized for the high gas inflow rate during quasar
activity (\S~\ref{s:smbh}), as well as for stellar mass black holes
that emit in the soft thermal state in LMXBs (\S~\ref{s:binary}).  To
better understand such objects, we start by exploring the structure of
thin disks that characterize the high accretion rate of quasars
\citep{1973A&A....24..337S,1973blho.conf..343N}.

We imagine a planar thin disk of cold gas orbiting a central black
hole.  Each gas element orbits at the local Keplerian velocity
$v_\phi=r\Omega =(GM/r)^{1/2}$ and spirals slowly inwards with
radial velocity $v_r\ll v_\phi$ as viscous torques transport its
angular momentum to the outer part of the disk. The associated viscous
stress generates heat, which is radiated away locally from the the
disk surface.  We assume that the disk is fed steadily and so it
manifests a constant mass accretion rate at all radii. Mass
conservation implies that
\begin{equation}
\dot{M}=2\pi r\Sigma v_r= {\rm constant}
\end{equation}
independent of radius and time, where $\Sigma(r)$ is the surface mass density of the disk.
The accretion rate $\dot{M}$ is a free parameter of the model, which is often
expressed with the accretion rate corresponding to the Eddington luminosity
as $\dot{m} = \dot{M}/\dot{M}_{\rm E}$. The corresponding angular
momentum flux is $\dot{M}r^2\Omega + {\rm constant}$.

In the limit of a geometrically thin disk with a scale height $H\ll r$,
the hydrodynamic equations decouple in the radial and vertical
directions.  In the radial direction,  the Keplerian
velocity profile introduces shear which dissipates heat as neighboring
fluid elements rub against each other.
This power provides a local radiative flux that leaves the system vertically from the top
and bottom surfaces of the disk,
\begin{equation}\label{e:Frad}
F_{\rm rad}=\frac{3}{8\pi}\dot{M}\Omega^2
\left(1 - \frac{r^2\Omega}{[r^{2}\Omega]_{\rm ISCO}}\right)
=\frac{3}{8\pi}\frac{G M \dot{M}}{r^3}
\left[1- \left({r_{\rm ISCO}\over r}\right)^{1/2}\right]\,.
\end{equation}
where we assumed that the torque vanishes at the innermost stable circular orbit (ISCO),
from where the gas plunges into the black hole over a free fall time. This sets the inner
boundary of the disk. Here $r_{\rm ISCO} = 6r_{\rm g}$ for a non-spinning black hole,
and $r_{\rm g}\leq r_{\rm ISCO} \leq 9r_{\rm g}$ for a spinning black hole, where $r_{\rm g}=GM/c^2$.
The total luminosity emitted by both faces of the disk is given by
\begin{equation}
L=\int_{r_{\rm ISCO}}^\infty 2F_{\rm rad}\times 2\pi r \D r
=\frac{1}{2}\frac{GM\dot{M}}{r_{\rm ISCO}} = \frac{r_g}{2 r_{\rm ISCO}}\dot{M} c^2,
\label{e:lumidisk}
\end{equation}
where we have ignored the radiation emitted inside the ISCO. This shows that the
radiative efficiency of the disk is $\epsilon = r_{\rm g}/(2 r_{\rm ISCO}) = 1/12 = 8.3\%$
for a nonspinning black hole and $50\%$ for a maximally spinning black hole in the
prograde direction. Note that this simple estimate neglected general-relativistic
corrections. A more detailed calculation gives similar results: $\epsilon=6\%$ for
nonspinning and $42\%$ for maximally spinning black holes \citep{Efficiency}.

In local thermodynamic equilibrium, the emergent flux from the surface
of the disk (equation \ref{e:Frad}) can be written in terms of the
disk surface temperature as $F_{\rm rad}= \sigmaSB T_{\rm d}^4$, where
$\sigmaSB$ is the Stephan-Boltzmann constant. This yields the
following radial profile for the surface temperature of the disk,
\begin{equation}
T_{\rm d} = \left(\frac{F_{\rm rad}}{\sigmaSB}\right)^{1/4}= 10^5~{\rm
K}~M_8^{-1/4}\dot{m}_{-1}^{1/4} r_1^{-3/4}
\left[1-\left({r\over r_{\rm ISCO}}\right)^{1/2}\right] .
\end{equation}
where $\dot{m}_{-1}=\dot{m}/0.1$, $r_1=r/(10r_{\rm g})$, and
$\dot{m}=\dot{M}/\dot{M}_{\rm Edd}$.  The corresponding thermal
blackbody spectrum peaks in the UV bands for quasars, and in the X-ray
band for stellar-mass black holes.  (Non-thermal X-ray emission from a
hot corona or a jet can supplement this disk emission.) Stellar-mass
black holes can therefore be important X-ray sources at high
redshifts, especially if they are incorporated in a binary system
where they accrete gas from a companion star (see \S~\ref{s:binary}).

Up to this point we did not need to adopt a model for the angular
momentum transport (or an effective viscosity) in the disk. Indeed,
the main observables, total luminosity and thermal spectrum, are very
robustly set for a radiatively efficient thin disk by three
parameters: $M$, $\dot{M}$, and the black hole spin which determines
$r_{\rm ISCO}$.  Other details, involving the disk thickness, the
radial surface density profile, and opacity however do depend on the
model of viscosity and the relative contribution of radiation to gas
pressure. The effective viscosity responsible is believed to be generated by
the magneto-rotational instability \citep{1998RvMP...70....1B}. 
Assuming that (i) the angular momentum flux is proportional
to the pressure in the disk with a dimensionless constant of
proportionality $\alpha$, (ii) the disk is very thin so that it can be
well approximated by vertically averaged quantities (one-zone model),
(iii) the disk is optically thick in the vertical direction so that
the gas and radiation are in thermal equilibrium, and (iv) the
self-gravity of the gas is negligible, one can algebraically express
all physical properties of the disk analytically
\citep{2004ApJ...608..108G}. The surface density and mid-plane
temperature are
\begin{eqnarray}\label{e:Sigma}
\Sigma &=& \frac{8\,(\mu \mproton/\kB)^{4/5}\sigmaSB^{1/5}}{3^{9/5}\,
\alpha^{4/5} \kappa^{1/5}} \frac{\beta^{(1-b)4/5}}{\Omega^{4/5}}
F_{\rm rad}^{3/5}\,,\\
T_{\rm c} &=& \frac{(\mu \mproton/\kB)^{1/5}\kappa^{1/5}}{(3\alpha\sigmaSB)^{1/5}}\frac{\beta^{(1-b)/5}}{\Omega^{1/5}}
F_{\rm rad}^{2/5}\,,
\label{e:Tc}
\end{eqnarray}
where $\beta = p_{\rm gas}/(p_{\rm gas}+p_{\rm rad})$ is the gas to total pressure given by
\begin{equation}\label{e:beta}
 \frac{\beta^{(1/2) + (b-1)/10}}{1-\beta} =
\frac{c\, [k /(\mu \mproton)]^{2/5} }{(3\,\alpha \sigmaSB)^{1/10} \kappa^{9/10}}\frac{ \Omega^{9/10}}{F_{\rm rad}^{4/5}}\,.
\end{equation}
Here $b=0$ and 1 denote models where viscosity is proportional to the
total pressure or only the gas pressure, respectively.
Equations~(\ref{e:Sigma}--\ref{e:beta}) along with
equation~(\ref{e:Frad}) define the disk model as a function of
radius. Equation~(\ref{e:beta}) shows that in the inner regions
$\beta\ll 1$, implying that radiation pressure dominates over gas
pressure.  In practice, radiation pressure dominates within
$600\,\alpha_1^{2/21} M_5^{2/21} r_{\rm g}$ if $\dot{M}=0.1
\dot{M}_{\rm Edd}$ with $\epsilon=10\%$ radiative efficiency.  The
opacity is dominated by electron scattering $\kappa_{\rm es}$ in the
inner region and free-free absorption in the outer regions
$\kappa_{\rm ff}$ (equation~\ref{e:kappa}).

Vertical hydrostatic equilibrium implies $H=c_s/\Omega$,
where $c_s=\sqrt{p/\rho}$ is the sound speed, where
$\rho=\Sigma/(2H)$ is the density,
$p=p_{\rm gas}+p_{\rm rad}$ is the pressure, $p_{\rm gas}=\rho \kB T_c/(\mu \mproton)$, and
$p_{\rm rad}=\frac{1}{3}a_{\rm rad}T_c^4$. Here $\mu$ is the mean molecular weight,
which is unity for a hydrogen plasma, and $\mu=0.615$ for $25\%$ helium and $75\%$ hydrogen
by mass. Combining these equations, leads to $H=\kappa F_{\rm rad}/[c \Omega^2 (1-\beta)]\sim(3\kappa/8\pi c)\dot{M}(1-\beta)^{-1}$.
When the accretion rate approaches the Eddington limit, $\beta\approx 0$,
and the disk bloats and $H$ approaches $r$ in the inner regions, violating the thin-disk assumption.
The model also breaks down if the accretion rate is several orders of magnitude below Eddington
because the disk becomes optically thin $\tau = \frac{1}{2}\kappa \Sigma < 1$.

The disk model described above ignores the self-gravity of the disk.
This assumption is inevitably violated at large radii, where the disk
becomes unstable to fragmentation due to its self-gravity, where the
Toomre parameter, $Q=(c_s\Omega/\pi G\Sigma)$, drops below unity.
Outside this radius, typically $\sim 2\times 10^4 r_g
(\alpha/0.1)^{0.6} (M/10^8\Msun)^{1.2}$, the gas in quasar accretion
disks would fragment into stars, and the stars may migrate inwards as
the gas accretes onto the black hole.  The energy output from stellar
winds and supernovae would supplement the viscous heating of the disk
and might regulate the disk to have $Q\sim 1$ outside the above
boundary.  We therefore conclude that star formation will inevitably
occur on larger scales, before the gas is driven into the accretion
disk that feeds the central black hole.  Indeed, the broad emission
lines of quasars display very high abundances of heavy elements in the
spectra out to arbitrarily high redshifts. Since the total amount of
mass in the disk interior to this radius makes only a small fraction
of the mass of the supermassive black hole, quasars must be fed by gas
that crosses this boundary after being vulnerable to fragmentation
\citep{2004ApJ...604L..45M}.

\subsection{Advection Dominated Accretion Flow} \label{s:adaf}

When the accretion rate is considerably lower than the Eddington limit
($\dot{m}\lesssim 10^{-2}$), the gas inflow switches to a
different mode, called a {\it Radiatively Inefficient Accretion Flow}
(RIAF) or an {\it Advection Dominated Accretion Flow} (ADAF), in which
either the cooling time or the photon diffusion time is much longer
than the accretion time of the gas and heat is mostly advected with
the gas into the black hole \citep{1994ApJ...428L..13N}.  At the low gas densities and high
temperatures characterizing this accretion mode, the Coulomb coupling
is weak and the electrons do not heat up to the proton temperature
even with the aid of plasma instabilities.  Viscosity heats primarily
the protons since they carry most of the momentum.  The other major
heat source, compression of the gas, also heats the protons more
effectively than the electrons.  As the gas falls inward and its density
$\rho$ rises, the temperature of each species $T$ increases
adiabatically as $T\propto \rho^{\gamma-1}$, where $\gamma$ is the
corresponding adiabatic index.  At radii $r\lesssim 10^2r_{\rm Sch}$,
the electrons are relativistic with $\gamma=4/3$ and so their
temperature rises inwards with increasing density as $T_e\propto
\rho^{1/3}$ while the protons are non-relativistic with $\gamma=5/3$
and so $T_{\rm p}\propto \rho^{2/3}$, yielding a two-temperature
plasma with the protons being much hotter than the electrons. Typical analytic
models \citep[and references therein]{1994ApJ...428L..13N,2008NewAR..51..733N}
yield $T_p\sim 10^{12}~{\rm K}(r/r_{\rm Sch})^{-1}, T_e\sim {\rm
min}(T_p,10^{9-11}~{\rm K})$. Since the typical sound speed is
comparable to the Keplerian speed at each radius, the geometry of the
flow is thick -- making RIAFs the viscous analogs of Bondi accretions.

Analytic models imply a radial velocity that is a factor of $\sim
\alpha$ smaller than the free-fall speed and an accretion time that is
a factor of $\sim \alpha$ longer than the free-fall time.  However,
since the sum of the kinetic and thermal energies of a proton is
comparable to its gravitational binding energy, RIAFs are expected to
be associated with strong outflows.

The radiative efficiency of RIAFs is smaller than the thin-disk value
($\epsilon$).  While the thin-disk value applies to high accretion
rates above some critical value, $\dot{m}>\dot{m}_{\rm crit}$, where
$\dot{m}$ is the accretion rate in Eddington units,
analytic RIAF models typically admit a radiative efficiency of
\begin{equation}
\epsilon \approx 10\,\dot{m}\quad{\rm if~~} \dot{m}\lesssim \dot{m}_{\rm crit}\,,
\end{equation}
with $\dot{m}_{\rm crit} \sim 0.01$.

The nature of the flow can change when the disk becomes radiatively inefficient, 
since thermal convection may compete with MHD turbulence in transporting energy 
and angular momentum.  In particular, the standard notion that viscosity is transported 
outwards in convective disks has been challenged in RIAFs 
\citep{2000ApJ...539..798N}. In such convection-dominated disks 
the density profile is altered considerably from the standard ADAF models.

Since at low redshifts mergers are rare and much of the gas in
galaxies has already been consumed in making stars, most local
supermassive black holes are characterized by a very low accretion
rate.  The resulting low luminosity of these dormant black holes, such
as the $4\times 10^6M_\odot$ black hole lurking at the center of the
Milky Way galaxy, is often described using RIAF/ADAF models.  Although
this mode of accretion is characterized by a low mass infall rate, it
could persist over a period of time that is orders of magnitude longer
than the quasar mode discussed earlier, so its contribution to the
growth of black holes in galactic nuclei may not be negligible.

\subsection{Circumbinary SMBH disks}
Galaxy mergers naturally lead to SMBH binaries, which can also produce
bright emission.  Direct X-ray imaging of active nuclei have revealed
several SMBH binaries at separations of $\sim$3\,kpc
\citep{2003ApJ...582L..15K,2004ApJ...600..634B,2008MNRAS.386..105B,2012ApJ...752...49C},
400\,pc \citep{2013arXiv1303.2630P}, and 150\,pc
\citep{2011Natur.477..431F}.  Additional evidence comes in the form of
a radio galaxy with a double core with a projected separation of
10\,pc \citep{2006ApJ...646...49R}, and several other observations of
radio galaxies, such as the wiggled shape of jets indicating
precession \citep{1993ApJ...409..130R}, the X-shaped morphologies of
radio lobes \citep{2002Sci...297.1310M,2004MNRAS.347.1357L}, the
interruption and recurrence of activity in double-double radio
galaxies \citep{2000MNRAS.315..371S,2003MNRAS.340..411L}, and the
elliptical motion of the unresolved core of 3C66B
\citep{2003Sci...300.1263S}, which have all been interpreted as
indirect evidence for SMBH binaries down to subparsec scales. Velocity
offsets in broad line spectra have also been observed in a handful of
AGN which may be attributed to the orbital motion of the SMBH binary
around the center of mass \citep{2012ApJS..201...23E,2013arXiv1306.4987J}.

The structure of the accretion flow around close SMBH binaries is less
understood.  Generally speaking, the dense nuclear gas around the
binary is expected to cool rapidly and settle in a rotationally
supported circumbinary disk \citep{2005ApJ...630..152E}.  If the
binary plane is initially misaligned with the disk, the gravitational
torques cause the disk to warp and twist. Viscosity then causes the
disk to align with the binary on a timescale $t_{\rm al} \sim
\max[\alpha (H/r)^{-2} t_{\rm orb}, (H/r)^{-1} t_{\rm orb} ]$ where $
t_{\rm orb}$ is the binary orbital period, $H$ is the scale-height,
and $\alpha$ is the dimensionless viscosity parameter of a thin disk
\citep[and see \S~\ref{s:disk}]{1999MNRAS.307...79I}.  For nearly
equal masses, the accretion flow may then develop a triple-disk
structure, forming individual accretion disks around the two binary
components, and a larger circumbinary disk feeding these smaller disks
\citep{2008ApJ...682.1134H}.  For unequal mass binaries $m/M\ll 0.1$,
the accretion disk geometry may be similar to a protoplanetary disk
with a planet orbiting in the disk. The secondary excites spiral
density waves in the disk which carry away energy and angular momentum
from the secondary.  This acts to reduce the density in a radial
annulus of comparable to the Roche lobe $R_R$ (equation
\ref{e:Roche}), regulates the inflow rate across the secondary orbit,
and leads to the inward migration of the secondary.  If the secondary
is capable of quenching the inflow efficiently, a hollow cavity may
form interior to its orbit with negligible accretion onto the primary.
The gradual accretion from the outer regions will make the gas pile up
outside the secondary's orbit similar to a river dam. The gas density
and pressure increases significantly, until the gas overflows in
non-axisymmetric streams, filling the cavity.

There are several interesting limiting cases for the circumbinary disk
structure. For extreme binary mass ratios, $q\lesssim 10^{-4}$, the
secondary cannot significantly modify the disk, which is then
described by a standard thin disk assuming that the accretion rate is
in the range $0.01\dot{m}<0.3$ (see \S~\ref{s:disk}). The spiral waves
generated in the disk by the secondary carry away angular momentum
very efficiently, shrinking the secondary's orbit quickly on the
so-called Type-1 migration timescale
\citep[$\sim10^4$--$10^5$\,yr,][]{1980ApJ...241..425G}.  For
comparable binary masses, $q\gtrsim 0.1$, a complete cavity forms, and
the secondary is driven inwards on the viscous timescale
($\sim10^7$\,yr) as the gas is flowing in from the outer regions
(Type-2 migration). However, typically the local disk mass is much
smaller than the secondary SMBH mass, implying that the gas piles up
before it can drive the binary in.  The accretion disk acquires a
self-similar structure well outside the secondary as long as there is
no inflow across the gap \citep{1999MNRAS.307...79I,2012arXiv1205.5017R}.  If the inflow
into the cavity is significant, the disk settles in a steady state
profile and the migration rate may be much slower \citep[Type-1.5
migration,][]{2012MNRAS.427.2660K}.

\begin{figure*}
\begin{center}
\includegraphics[width=6cm]{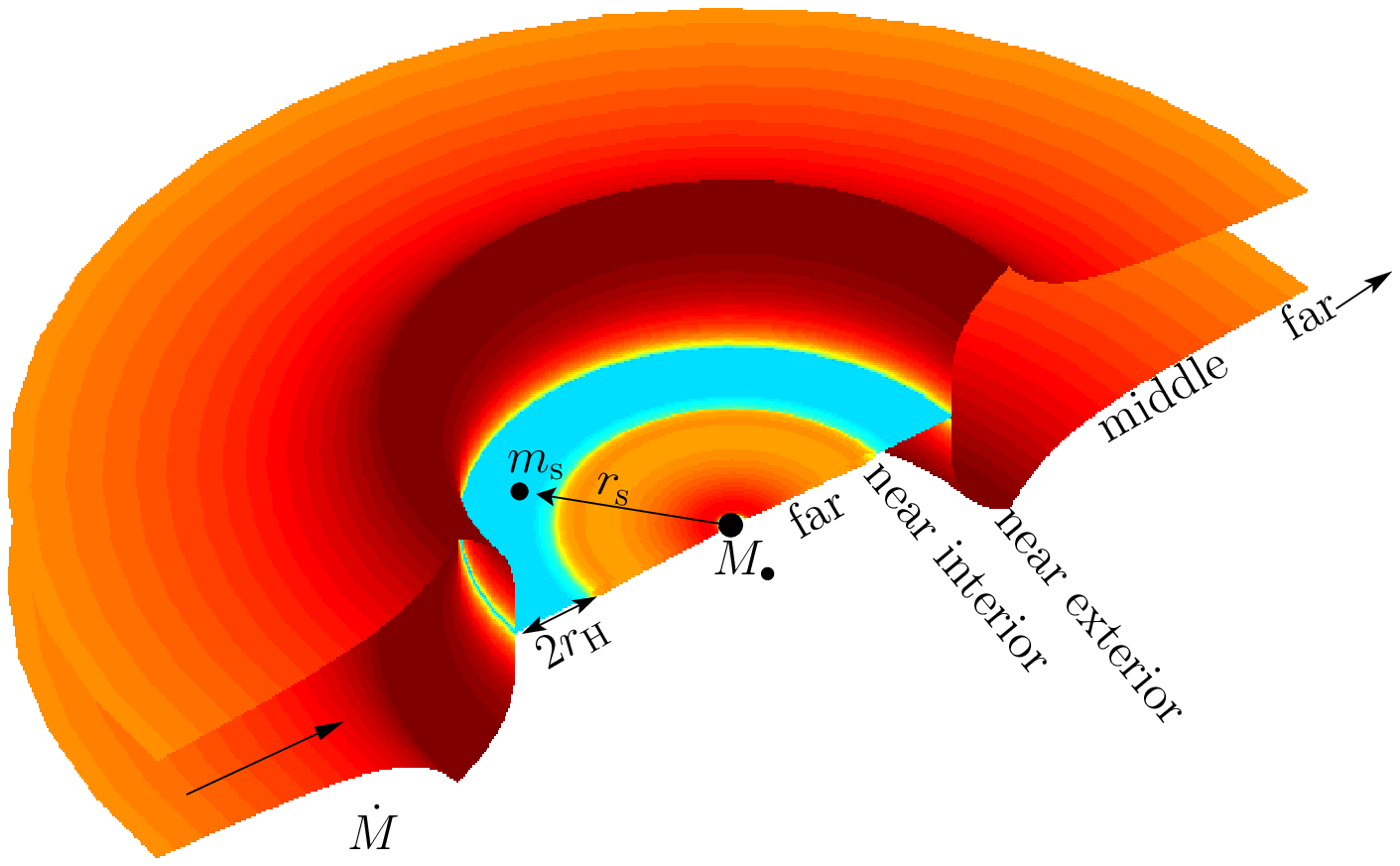}
\includegraphics[width=6cm]{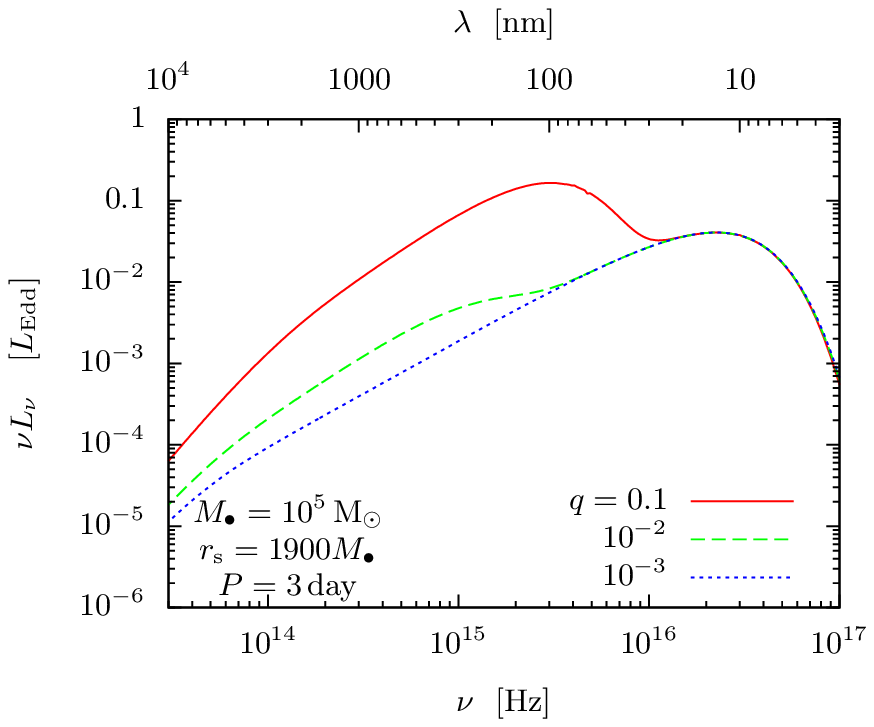}
\caption{
Gas pile up due to a secondary in the accretion disk (left) and the corresponding thermal continuum
spectrum for mass ratios $q=0.1$, 0.01, and 0.001 (right).
 Figure credit: \citet{2012MNRAS.427.2680K}. }
\label{f:circumbinary}
\end{center}
\end{figure*}

The accretion disk luminosity depends on whether the inner disk is
truncated by tidal effects of the secondary and depends on the amount
of gas pile up outside the secondary's orbit. The radiative efficiency
may be larger than equation~(\ref{e:lumidisk}) due to the energy
release corresponding to the inward migration of the secondary. This
excess power comes from the part of the accretion disk outside the
secondary's orbit where the gas pile up is significant.
Figure~\ref{f:circumbinary} shows that this can make the disk
significantly brighter in the optical bands for $q<0.01$
\citep{2012MNRAS.427.2680K}.  Furthermore, hydrodynamical simulations show that the
accretion flow into the inner regions is expected to display periodic
variability on roughly the orbital timescale. The variability power
spectrum has discrete peaks, where the frequencies and harmonic
weights depend on the binary mass ratio, accretion rate, and the
$\alpha$ parameter
\citep{2008ApJ...672...83M,2009MNRAS.393.1423C,2012ApJ...749..118S,2012ApJ...755...51N,2012arXiv1210.0536D}.
However, further improvements are needed to include radiation pressure
and extended to cover several viscous timescales to make more
quantitative predictions on circumbinary accretion.

Supermassive black hole binaries may be identified in future surveys
based on the velocity offsets in their broad line spectra
\citep{2010ApJ...725..249S,2012ApJS..201...23E}, the expected periodic
variability of the accretion luminosity on weeks-to-months timescales
\citep{2009ApJ...700.1952H}, the excess brightness in the optical
bands of the thermal spectrum corresponding to the pile up
\citep{2012MNRAS.427.2680K}, the variability of relativistic iron
lines due to the orbital motion of the secondary and the effects of a
gap in the disk \citep{McKernan13}, or through the gravitational waves
emitted by the binary \citep{2008ApJ...684..870K,2011PhRvD..84b2002L}.

\section{Feeding Black Holes with Stars}

\subsection{Tidal Disruption Events}
Most supermassive black holes in the local Universe are not accreting
gas. They are typically lurking at the centers of galaxies, surrounded
by a dense population of stars.  A star that comes close to the black
hole is torn apart by tidal gravity.  The spaghetti-like stellar
debris falling back onto the black hole forms an accretion disk and
produces a luminous electromagnetic transient lasting weeks to years,
observable from cosmological distances \citep{1976MNRAS.176..633F}.
Around $25$ tidal disruption events have been observed to date in
X-ray, UV, optical, and radio bands \citep[and references
therein]{2012EPJWC..3902001K,2012EPJWC..3903001G}. These observations
imply a stellar disruption rate of $3\times 10^{-5}\,{\rm
yr}^{-1}\,{\rm galaxy}^{-1}$ \citep{2012EPJWC..3908002V}, consistent
with theoretical rate estimates
\citep{1988Natur.333..523R,2004ApJ...600..149W}.

The characteristic light-curve produced by tidal disruption events is derived
as follows. The maximum distance at which the tidal field of the black hole
is capable of unbinding a star is the tidal radius,
\begin{equation}\label{e:tidal}
 R_t = \left(\frac{M_{\rm BH}}{M_*}\right)^{1/3} R_* =
 46\, r_{\rm g} \left(\frac{M_{\rm BH}}{10^6\,\Msun} \right)^{-2/3} \left(\frac{M_*}{M_{\odot}}\right)^{-1/3} \frac{R_*}{R_{\odot}}\,.
\end{equation}
The disruption occurs outside of the black hole horizon for $M_{\rm BH}<10^8\,\Msun$ for main
sequence stars, $\lesssim 10^6\,\Msun$ for white dwarfs and $\lesssim 5\,\Msun$
for neutron stars \citep[see][for highly
spinning black holes where these bounds may be a factor 10 larger]{2012PhRvD..85b4037K}.
After tidal disruption, approximately half of the stellar mass is bound to the black hole and half is unbound.
The most bound material returns to pericenter after a time
\begin{equation}
 t_{\rm fb} = \frac{2\pi}{6^{3/2}} \left(\frac{r_p}{R_{*}}\right)^{3/2} t_{p}
= 20 \left(\frac{M_{\rm BH}}{10^6\,\Msun} \right)^{3/2} \left(\frac{r_p}{6r_g}\right)^{3} \left(\frac{R_*}{\Rsun}\right)^{-3/2}\, \rm min
\end{equation}
where $R_p$ is the pericenter distance,
$t_p=(GM_{\rm BH}/R_p^3)^{-1/2}$ is the pericenter timescale.
The fallback rate of bound material is then \citep{1988Natur.333..523R}
\begin{equation}
  \dot{M}_{\rm fb} = \frac{\D M}{\D E} \frac{\D E}{\D t} = \frac{1}{3} \frac{M_{*}}{t_{\rm fb}} \left(\frac{t}{t_{\rm fb}}\right)^{-5/3},
\end{equation}
where $E\sim G M_{\rm BH}/a$ is the orbital energy,
and $a\propto t^{-3/2}$ from Kepler's law, and we assumed that $\D M/\D E$ is constant.
The timescale for the light-curve to reach the $t^{-5/3}$ limit depends on the
stellar structure and spin through $\D M/\D E$  \citep{2009MNRAS.392..332L,2012ApJ...757..134M,2013ApJ...767...25G,2012arXiv1210.3374S}.
This Newtonian estimate is modified for smaller stellar orbital eccentricities
and by GR corrections, which depend on the spin of the black hole
\citep{2012PhRvD..86f4026K,2012arXiv1210.1333H,2013arXiv1303.4837D}.

Once the infalling gas returns to pericenter, it is expected to shock
and form an accretion disk around the black hole.  Initially, weeks to
months after the disruption, the fallback rate is very
super-Eddington. This phase is probably described by an advective slim
disk with powerful outflows
\citep{2009MNRAS.400.2070S,2012ApJ...760..103D,2013arXiv1301.1982T}.
Months to a year later, the fallback rate becomes sub-Eddington, and
the disk may radiatively cool efficiently to form a thin disk
(\S~\ref{s:disk}).  Finally, several years after the disruption, the
fallback rate becomes very sub-Eddington $\dot{M}_{\rm fb}\lesssim
10^{-2}\dot{M}_{\rm Edd}$, and the disk becomes radiatively
inefficient and hot (\S~\ref{s:adaf}) which may be accompanied by a
jet.  Two possible tidal disruption event sources have been observed
to display jet activity \citep[Swift J1644+57 and Swift
J2058.4+0516,][]{2011Natur.476..421B,2012ApJ...753...77C}.  The lack
of evidence for Lense-Thirring precession of the jet away from the
observer suggests that the jet is aligned with the SMBH spin and not
the accretion disk in these events \citep{2012PhRvL.108f1302S}.

The bolometric luminosity of the system is $L = \epsilon \dot{M}_{\rm
fb} c^2 \propto t^{-5/3}$.  In the thin disk phase, the multicolor
blackbody spectrum of the accretion disk peaks in the soft X-ray
bands.  The light-curve in the X-ray bands follows $t^{-5/3}$ for
several years, indeed consistent with many observed sources
\citep{2012EPJWC..3902001K}.  Optical and UV wavelengths initially
fall on the Rayleigh-Jeans tail of the disk spectrum, and the
luminosity is initially proportional to the temperature
$L_{\nu}\propto T_{\nu} \propto \dot{M}_{\rm fb}^{1/4}\propto
t^{-5/12}$, and eventually converges to the $t^{-5/3}$ profile as the
disk cools \citep{2011MNRAS.410..359L}.  Such a shallower decline has
been observed in the optical and UV bands for Swift J2058
\citep{2012ApJ...753...77C}.

Upcoming surveys will detect more hundreds to thousands of tidal
disruption events (TDEs). Multi-wavelength observations will enable us
to probe the supermassive and intermediate-mass black hole
populations, allowing us to measure their masses and spins
\citep{2012PhRvD..86f4026K}.  Supermassive black hole binaries may
imply a transient very high TDE rate of $1\,{\rm yr}^{-1}$ and account
overall for $10\%$ of the cosmic TDE rate \citep{2011ApJ...729...13C}.
Detecting multiple TDEs in the same galaxy within a few years time
could be a signpost for a MBH binary. Merging SMBH receive a
gravitational wave recoil kick due to anisotropic gravitational
radiation. The recoiling black hole could produce tidal disruption
flares spatially offset from the galactic center, which might
contribute $1\%$ of the TDE rates \citep{2012MNRAS.422.1933S}.
Finally, white dwarfs passing close to IMBHs may be detonated by the
tidal gravity which would appear as underluminous supernovae
\citep{1982Natur.296..211C,2009ApJ...695..404R,2012ApJ...749..117H}.

\subsection{Fueling Active Galactic Nuclei Accretion Disks with Stars}
In the standard theory of AGN accretion, the effective viscosity responsible
for delivering material onto the black hole is believed to be generated by
the magneto-rotational instability (\S~\ref{s:disk}). An interesting
alternative idea was examined by \citet{2001ApJ...552..793G} in which
massive objects embedded in the accretion disk serve as an effective
source of viscosity. These objects drive density waves
in the disk which transport energy and angular momentum like a kinematic viscosity.
In particular, stars in the stellar cluster around the supermassive black hole that cross the disk
become captured in the disk due to accretion and hydrodynamical drag.
\citet{2005ApJ...619...30M} have argued that the matter of these stars
is sufficient to replenish the disk and fuel AGN. This process can also explain the
observed correlation between the black hole mass and velocity dispersion
of the stellar cluster ($M$-$\sigma$ relation), although other explanations based
on AGN feedback are also possible (see Chapter 5.6 in this book).
Additionally, stars may also form in the outer regions of accretion disks
due to gravitational fragmentation, and migrate inwards as they
interact with the disk to fuel the AGN \citep{2007MNRAS.374..515L}.

The fate of the stars embedded in accretion disks is not well understood.
Once they approach the black hole within the tidal radius (equation~\ref{e:tidal}),
they are tidally disrupted. However, their structure might change considerably
before reaching this radius if they migrate across a gaseous disk.
Accretion from the disk and mergers with other stars can lead to the formation of a
supermassive star or an intermediate mass black hole already
at large distances from the black hole \citep{2004ApJ...608..108G,2012MNRAS.425..460M}.
If so, these objects orbiting around supermassive black holes will generate
gravitational waves, and provide a sensitive new probe of the accretion disk
\citep{2011PhRvD..84b4032K}.


\nocite{*}
\bibliographystyle{aps-nameyear}      
\bibliography{p}   

\end{document}